\documentclass{llncs}

\usepackage[np]{numprint}
\usepackage[utf8]{inputenc}
\usepackage[T1]{fontenc}
\usepackage{graphicx}
\usepackage{arydshln}
\usepackage{booktabs}
\usepackage{caption}
\usepackage[caption=false]{subfig}
\usepackage{arydshln}
\usepackage{multirow}
\usepackage{import}
\usepackage{amssymb}
\usepackage{fp}
\usepackage{comment}
\usepackage{balance}
\usepackage[breaklinks]{hyperref}
\usepackage[sort,nocompress]{cite}
\usepackage{cleveref}
\usepackage{xspace}
\usepackage[binary-units]{siunitx}
\DeclareSIUnit{\packet}{p}
\usepackage[nolist]{acronym}

\usepackage[disable]{todonotes} %
\presetkeys{todonotes}{fancyline, color=olive!30, inline}{}

\newcommand{\todoam}[1]{\todo[color=red!40]{\textbf{@Aniss: } #1}}
\newcommand{\new}[1]{\textcolor{blue}\relax{#1}} %
\newcommand{\budget}[1]{\todo[color=yellow!40]{\textit{Budget: #1 page(s)}}}
\renewcommand{\budget}[1]{}
\usepackage{pifont} %
\usepackage[normalem]{ulem}
\usepackage{microtype}

\usepackage{blindtext}

\usepackage{floatrow}
\newfloatcommand{capbtabbox}{table}[][\FBwidth]

\newcommand{\zmap}{ZMap\xspace}
\newcommand{\zmapsix}{ZMapv6\xspace}
\newcommand{\yarrp}{Yarrp\xspace}

\newcommand{\eg}{e.g.,\xspace}
\newcommand{\ie}{i.e.,\xspace}
\newcommand{\etal}{et al.\xspace}
\newcommand{\ixp}{large European IXP\xspace}

\newcommand{\one}{(1)\xspace}
\newcommand{\two}{(2)\xspace}

\newif\ifcutoptional
\cutoptionalfalse
\DeclareSIUnit{\nothing}{\relax}

\newcommand{\mn}{\mega\nothing}
\newcommand{\kn}{\kilo\nothing}

\newcommand{\sk}[1]{\SI{#1}{\kn}}
\newcommand{\sm}[1]{\SI{#1}{\mn}}

\newcommand{\sperc}[1]{\SI{#1}{\percent}}

\renewcommand*{\paragraph}[1]{%
    \vspace{.5em}
    \noindent
    {\normalfont \bf #1}
}

\usepackage[absolute,showboxes]{textpos}

\begin{document}

\title{Zeroing in on Port 0 Traffic in the Wild}

\author{Aniss Maghsoudlou \and
Oliver Gasser \and
Anja Feldmann}
\institute{Max Planck Institute for Informatics\\
\email{\{aniss,oliver.gasser,anja\}@mpi-inf.mpg.de}
}

\setlength{\TPHorizModule}{\paperwidth}
\setlength{\TPVertModule}{\paperheight}
\TPMargin{5pt}
\begin{textblock}{0.8}(0.1,0.02)
    \noindent
    \footnotesize
    If you cite this paper, please use the PAM '21 reference:
    Aniss Maghsoudlou, Oliver Gasser, Anja Feldmann.
    Zeroing in on Port 0 Traffic in the Wild.
    In \textit{Passive and Active Measurement Conference 2021 (PAM '21), March 29--31, 2021, Virtual Event.}
    Springer Nature Switzerland AG 2021, PAM 2021, LNCS 12671, pp. 1--17, 2021.
    \url{https://doi.org/10.1007/978-3-030-72582-2_32}
\end{textblock}

\maketitle

\begin{abstract}
\budget{0.4}

Internet services leverage transport protocol port numbers to specify the source and destination application layer protocols.
While using port 0 is not allowed in most transport protocols, we see a non-negligible share of traffic using port 0 in the Internet.

In this study, we dissect port 0 traffic to infer its possible origins and causes using five complementing flow-level and packet-level datasets.
We observe 73 GB of port 0 traffic in one week of IXP traffic, most of which we identify as an artifact of packet fragmentation.
In our packet-level datasets, most traffic is originated from a small number of hosts and while most of the packets have no payload, a major fraction of packets containing payload belong to the BitTorrent protocol. 
Moreover, we find unique traffic patterns commonly seen in scanning.
In addition to analyzing passive traces, we also conduct an active measurement campaign to study how different networks react to port 0 traffic.
We find an unexpectedly high response rate for TCP port 0 probes in IPv4, with very low response rates with other protocol types.
Finally, we will be running continuous port 0 measurements and providing the results to the measurement community.

\end{abstract}

\section{Introduction} \label{section:introduction}
\budget{0.6}

Transport protocols use port numbers to identify different Internet services.
Common port numbers are TCP/80 and TCP/443 for the Web, TCP/25 for SMTP, or UDP/443 for QUIC.
There are different categories of port numbers: Officially registered ports at IANA \cite{ianaports}, unofficially but well-known ports, and dynamic ports, which cannot be registered and are free to use by anyone.
In contrast, there are also some ports which are reserved and should not be used.
One of these reserved port numbers is port 0.
It is reserved in most common transport layer protocols, \ie TCP~\cite{rfc870}, UDP~\cite{rfc870}, UDP-Lite~\cite{rfc3828}, and SCTP~\cite{rfc4960}.
When providing a port number 0 to the \texttt{bind()} system call to establish a connection, operating systems generally choose a free port from the dynamic range \cite{port0linux,port0windows}.
Therefore, one needs to create a \new{raw} socket in order to send port 0 packets.
However, previous work has shown that there is a non-negligible share of traffic using port number 0 \new{both in darknets and the Internet}\cite{bou2014multidimensional,luchs2019curious,maghsoudlou2020reserved}.

In this work, we shed light on port 0 traffic in the Internet, by analyzing the traffic from real networks, rather than darknets as is done in most related work, and by performing active measurements to survey the real-world reaction of hosts and routers to port 0 traffic.

To the best of our knowledge, this is the first work which conducts both active and passive measurements on port 0 in the Internet, to better understand port 0 traffic characteristics and origins.
Specifically, this work has the following three main contributions:
\begin{itemize}
    \item We leverage a flow-level dataset from a \ixp to inspect the origins of port 0 traffic (cf. \Cref{section:flow-level}). 
We find that out of the top 10 ASes originating port 0 traffic, the majority does not follow typical diurnal patterns of common protocols such as TCP/80. %
    \item We inspect four packet-level datasets to discover the actual contents and detailed characteristics of port 0 packets (cf. \Cref{section:packet-level}). We show that the majority of non-empty packets in UDP are related to BitTorrent. We find that most TCP packets do not contain any payload and are one-way. However, most of the two-way TCP streams are scanning artifacts. %
    \item We perform active measurements \new{both in IPv4 and IPv6} to gain a tangible perspective over port 0 responsive IP addresses (cf. \Cref{section:active-measurement}). We find that IPv4 traffic using TCP uncovers a substantial number of responsive hosts in a small number of ASes. We also perform traceroute-style active measurements to better understand port 0 traffic filtering in wild, and find discrepancies between IPv4 and IPv6. Finally, we will run periodic port 0 measurements and make the results available to the research community.
\end{itemize}

\section{Related Work}
\label{section:related-work}
\budget{0.5}

Already in 1983, Reynolds and Postel specified that port number 0 is reserved in TCP and UDP~\cite{rfc870}.
Over the course of several years, similar provisions have been introduced for other transport protocols as well~\cite{rfc3828,rfc4960}.
Traffic sent from or to port 0 thus violates these specifications.
Fittingly, most reports on port 0 traffic are associated with DDoS attacks~\cite{cloudflare2017amplification,isc,endace} and malformed packets \cite{bykova2002malformed}.

Even though there is traffic on port 0 in the Internet, there is little research on its root causes.
Motivated by port 0 traffic spikes observed in November 2013 at the Internet Storm Center and reports from security researchers at Cisco Systems, Bou-Harb \etal~\cite{bou2014multidimensional} study port 0 traffic on 30 GB of darknet data.
They filter out any misconfigured traffic and packets with non-conforming TCP flags common in backscatter traffic \cite{wustrow2010backscatter}.
Using fingerprinting techniques~\cite{bou2014fingerprinting}, they argued that more than 97\% of their identified port 0 traffic was related to probing activities, some orchestrated by malware.

In 2019, Luchs and Doerr \cite{luchs2019curious} revisit the case of port 0 traffic, by studying data obtained from a /15 darknet over a period of three years.
They find that out of about \num{33000} source IP addresses involved in port 0 traffic,  10\% can be attributed to DDoS attacks, 6\% to OS fingerprinting, and less than 1\% to scanning activities.
When aggregating by the number of packets instead, scanning traffic dominates with 48\% of all port 0 packets.

More recently, Maghsoudlou \etal \cite{maghsoudlou2020reserved} analyze port 0 traffic for a single passive measurement source.
Similarly to our results, they find that a small number of ASes are responsible for about half of all port 0 traffic.

In contrast to the related work \cite{bou2014multidimensional,luchs2019curious,maghsoudlou2020reserved}, which all focus their efforts on the analysis of a single passive data source, in this paper we analyze four complementing passive datasets in addition to conducting an active measurement campaign to better understand port 0 traffic in the wild.

\section{Datasets Overview} \label{section:datasets}
\budget{1}

We leverage two different kinds of passive datasets to study port 0 traffic characteristics: Flow-level and packet-level data.
\new{Throughout the paper, port 0 traffic refers to the subset of the traffic which has either source port or destination port or both set to zero.}
Flow-level data gives us a high-level overview of Internet traffic and can be used to analyze the aggregate flow of traffic.
In our case, we use one week of IPFIX flow data from a \ixp. %
On the other hand, to be able to dissect detailed traffic characteristics like fragmentation, header flags, and different payloads, we need to inspect every single packet. 
Therefore, we use four different packet-level datasets, namely long-term and short-term MAWI, CAIDA, and Waikato. 
Different packet-level datasets are used to cover different geographical and temporal vantage points.

As shown in Table \ref{tab:datasets}, we use the following datasets:
\begin{description}
    \item[IXP] One week of sampled IPFIX data from the end of January 2020 captured at a \ixp. %
    \item[MAWI] These datasets \cite{mawi} contain packet traces from the transit link of the WIDE backbone \cite{wide} to the upstream ISP \new{captured at samplepoint-F}.
        They include partial packet payload.
        To obtain a more comprehensive view, we use two variants of MAWI datasets:
    \subitem \textbf{MAWI-long} This dataset captures 15-minute snapshots each month from January 2007 to July 2020.
    \subitem \textbf{MAWI-short} We also use the most recent MAWI dataset being part of the Day in the Life of the Internet project \cite{ditl}, which is April 8--9, 2020.
    \item[CAIDA] This dataset \cite{caidapassive} contains anonymized packet traces without payload from CAIDA's passive monitors.
        For our analysis we use the most recent dataset available at the time of writing, which is the one-hour period from 14:00--15:00 UTC recorded on January 17, 2019.
    \item[Waikato] This dataset \cite{waikato} contains packet header traces including the first few bytes of payload and is captured at the border of the University of Waikato network in New Zealand.
\end{description}

\begin{table}[t]
\resizebox{\columnwidth}{!}{
	\begin{tabular}{ l@{\hskip 0.3cm}l@{\hskip 0.3cm}l@{\hskip 0.3cm}l@{\hskip 0.3cm}l@{\hskip 0.3cm}l }
        \toprule
		Dataset           & IXP              & MAWI-long     & MAWI-short          & Waikato         & CAIDA\\ %
        \midrule
		Timespan          & Jan. 25--31, 2020 & 2006--2020     & Apr. 8--9, 2020    & Apr.--Nov., 2011    & Jan. 17, 2019\\
		Duration          & 1 week           & 14 years      & 2 days        & 86 Days              & 2 hours\\
        Format            & Flows            & Packets       & Packets       & Packets              & Packets \\
        \%IPv4,IPv6 (Port0) & 99.8\%,0.2\%     & 100\%,0\%   & 100\%,0\%         & 100\%,0\%                 & 99.7\%,0.3\% \\
        \%UDP,TCP (Port0) & 96.8\%,3.2\%     & 22.4\%,77.6\%   & 30.2\%,69.8\%         & 15.5\%,84.5\%         & 43.8\%,56.2\% \\
		Payload           & No               & Yes           & Yes           & Yes                  & No\\
		Sampled           & Packet-based      & Time-based           & No            & No                   & No \\
		\# Packets        & $34.3 \times 10^9$  & $23 \times 10^{9}$   & $15.9 \times 10^9$  & $27.822 \times 10^9$    & $8.2 \times 10^9$ \\
		\% Port 0 packets & 0.25             & 0.0008         & 0.0001       & 0.002                & 0.0002 \\
		\# Bytes          & 25.5 TB           & 14.6 TB        & 6.7 TB     & 16.9 TB             & 4.3 TB \\
		\% Port 0 bytes   & 0.28              & 0.00012        & 0.0002     &0.001             & 0.00002 \\
        \bottomrule
	\end{tabular}\textsc{}
}
\caption{Overview of passive port 0 datasets.}
\label{tab:datasets}
\end{table}

We analyze port 0 traffic seen in passive data in detail in \Cref{section:flow-level,section:packet-level}.
In addition to passive flow and packet data, we also conduct active measurements.
More specifically, we run two types of measurements to analyze responsiveness on port 0 and filtering of port 0 traffic in the Internet:

\begin{description}
    \item[Port scan] We use \zmap \cite{durumeric2013zmap,zmap} and \zmapsix \cite{zmapv6} to find responsive addresses on port 0.
        In IPv4 we conduct Internet-wide measurements, in IPv6 we leverage an IPv6 hitlist \cite{gasser2016scanning,gasser2018clusters,hlv2}.
    \item[Traceroute] We use \yarrp \cite{beverly2016yarrp,yarrp} to traceroute addresses in IPv4 and IPv6 prefixes in order to analyze port 0 traffic filtering in the Internet.
\end{description}

We present results from our active measurement campaign in \Cref{section:active-measurement}.
By leveraging both passive and active measurements we can analyze different aspects of port 0 traffic in the wild.

\subsection{Ethical Considerations}

Before conducting active measurements, we follow an internal multi-party approval process which incorporates proposals by Partridge and Allman \cite{partridge2016ethical} and Dittrich \etal \cite{dittrich2012menlo}.
We follow scanning best practices \cite{durumeric2013zmap} by limiting our probing rate, maintaining a blocklist, and using dedicated servers with informing rDNS names, websites, and abuse contacts.
During our active measurements, we received one email asking to be blocked, to which we immediately complied.

When analyzing passive flow and packet data, we fully comply with the respective NDAs and do not share any personally identifiable information.
Contrary to the active measurements, we will not publish any passive measurement data.

\subsection{Reproducible Research} \label{section:reproducible-research}

To foster reproducibility in measurement research \cite{AcmArtifacts,reproduc2017}, \new{we make data, source code, and analysis tools of our active measurements publicly available \cite{dataset}}.
Due to privacy reasons we will not publish data from the passive datasets.

\subsection{Continuous Port 0 Measurements}

To allow further analysis of port 0 responsiveness and filtering over time, we \new{periodically run} active port 0 measurements.
The raw results of these measurements \new{are} publicly available for fellow researchers at:

\begin{center}
    \new{\href{https://inet-port0.mpi-inf.mpg.de/}{\textbf{inet-port0.mpi-inf.mpg.de}}}
\end{center}

\section{Flow-level Analysis} \label{section:flow-level}
\budget{1}
Analyzing the traffic flowing between different Autonomous Systems is helpful to detect high-level patterns.
To investigate port 0 traffic patterns, we use the IXP dataset and inspect the ASes originating or being targeted by port 0 traffic.
In one week of IXP flow data, we find \num{23000} ASes contributing to port 0 traffic.
We observe that the source AS with highest number of packets in sends port 0 traffic to 4357 distinct destination ASes. Also, the destination AS with highest number of port 0 packets being destined to, is targeted by 1245 distinct source ASes.

We also observe that in 9 out of 10 top source ASes involved in port 0 traffic, port number 0 is among the top-5 source and destination port numbers along with TCP/80 (HTTP) and TCP/443 (HTTPS).
\new{We find that more than 99\% of port 0 traffic has both source and destination port set to zero.}
Interestingly, more than 99\% of all TCP traffic contains no TCP flags. 
This leads us to believe that this is not actual port 0 traffic and is most likely an artifact of packet fragmentation \cite{kopp2021ddos}, which is incorrectly classified as TCP/0 traffic by the flow exporter \cite{nokiaswitch}.
\new{We also analyze the 1\% of the TCP traffic with non-zero TCP flags, composed of 867 packets. We find that 30\% of this traffic sets their TCP flags to CWR/URG/ACK, 27\% to ACK only, and 25\% to URG/ACK/PSH/SYN.
62\% of this traffic has an average packet size of less than 100 bytes, while 18\% has an average packet size of more than 1480 bytes.}  
To investigate more in-depth on how different networks react to port 0 traffic, we perform active measurements (cf. \Cref{section:active-measurement}).

\begin{figure}[htb]
    \captionsetup{skip=.25em}
    \centering
    \includegraphics[width=0.5\columnwidth]{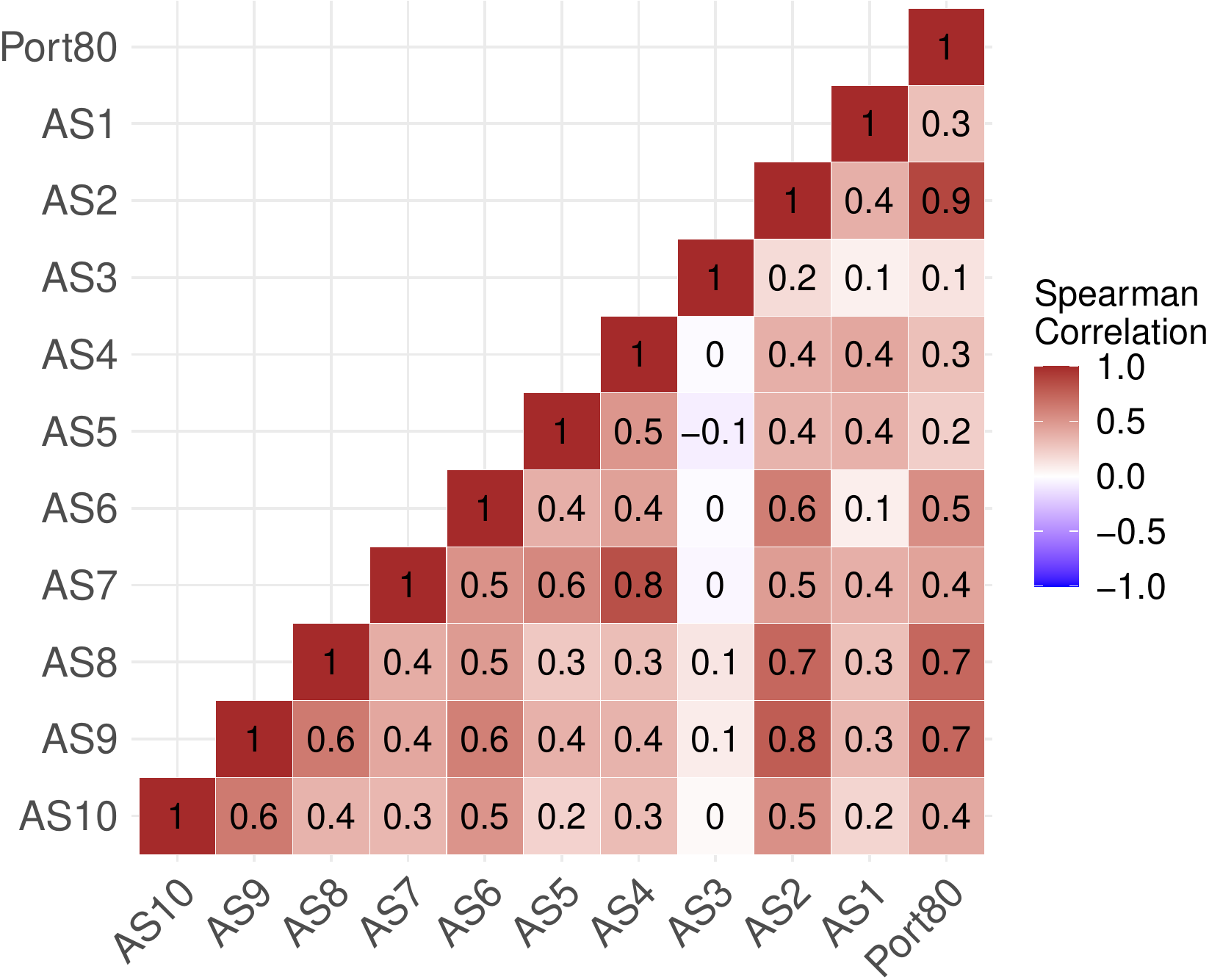}
    \caption{Correlation coefficients between port 80 traffic and the top 10 source ASes involved in port 0 traffic in the IXP.}
    \label{fig:correlation}
\end{figure}

To further investigate origins and causes of port 0 traffic, we analyze the diurnal patterns of traffic originated by the top 10 source ASes and compare them with the more common Web traffic on TCP/80.
\Cref{fig:correlation} shows a heatmap of the Spearman correlation of the diurnal patterns of these ASes and TCP/80 traffic.
We see that while AS2 is the most correlated to TCP/80 traffic, AS4 and AS7 show highly similar patterns to each other and moderate correlation to TCP/80 traffic. 
Moreover, AS3 shows a unique pattern with no correlation to either other ASes or TCP/80.\\
AS3 is a cloud computing provider while other ASes are web hosting providers, ISPs, or telecommunication companies.
The unique traffic pattern originated by AS3 implies irregular usage such as scanning or reset attack. 
For the interested reader we provide a time series plot for the aforementioned ASes in \Cref{sec:appendix:passive}.

To better understand the causes of port 0 traffic, we analyze average payload sizes observed in the IXP dataset.
For easier comparison with the packet-level datasets (cf. \Cref{section:packet-level}), we choose to analyze the payload size instead of the average packet size reported directly in the flow data.
We estimate the payload size by subtracting the IP and TCP/UDP headers without options. 
As shown in \Cref{fig:cdf-payload}, for TCP, we observe that nearly 88\% of packets are smaller than 100 bytes, while in UDP, more than 75\% of packets are larger than 100 bytes. 
Having roughly 20\% full-sized packets in UDP, along with many mid-sized packets, indicates possible fragmentation. 
Unfortunately, our IPFIX dataset does not include fragmentation information for IPv4 flows. It does, however, include information about the IPv6 next header value. 
We find no IPv6 flows with the next header value set to fragmentation (\ie 44). 
To investigate further on the exact fragmentation header flag, we inspect the IPFIX field containing a list of all IPv6 extension headers in a flow. We find, however, that the content of this IPFIX field does not conform to the IPFIX specifications as defined by the RFC.
This is possibly due to an erroneous early version of the RFC, which has since been corrected~\cite{erratum1738}.
As IPFIX datasets usually depend heavily on how their exporter is implemented, researchers who would like to work on them should be extra cautious to make sure that their data is flawless.

To summarize, multiple indicators lead us to believe that most of port 0 traffic seen at the IXP is an artifact of packet fragmentation. 
\new{Nevertheless, we find that the IXP data gives valuable information on diurnal patterns.}
\new{By} analyzing the correlation between diurnal patterns of different ASes and port 80 traffic, we find one AS deviating heavily from the common diurnal patterns. This indicates possible scanning or other irregular activities which requires a more in-depth analysis which can only be performed on packet-level data.
Therefore, we analyze the four packet-level datasets in the upcoming section.
\section{Packet-level Analysis} \label{section:packet-level}
\budget{1}

\begin{figure}[b!]
	\captionsetup{skip=.25em}
	\centering
	\includegraphics[width=\columnwidth]{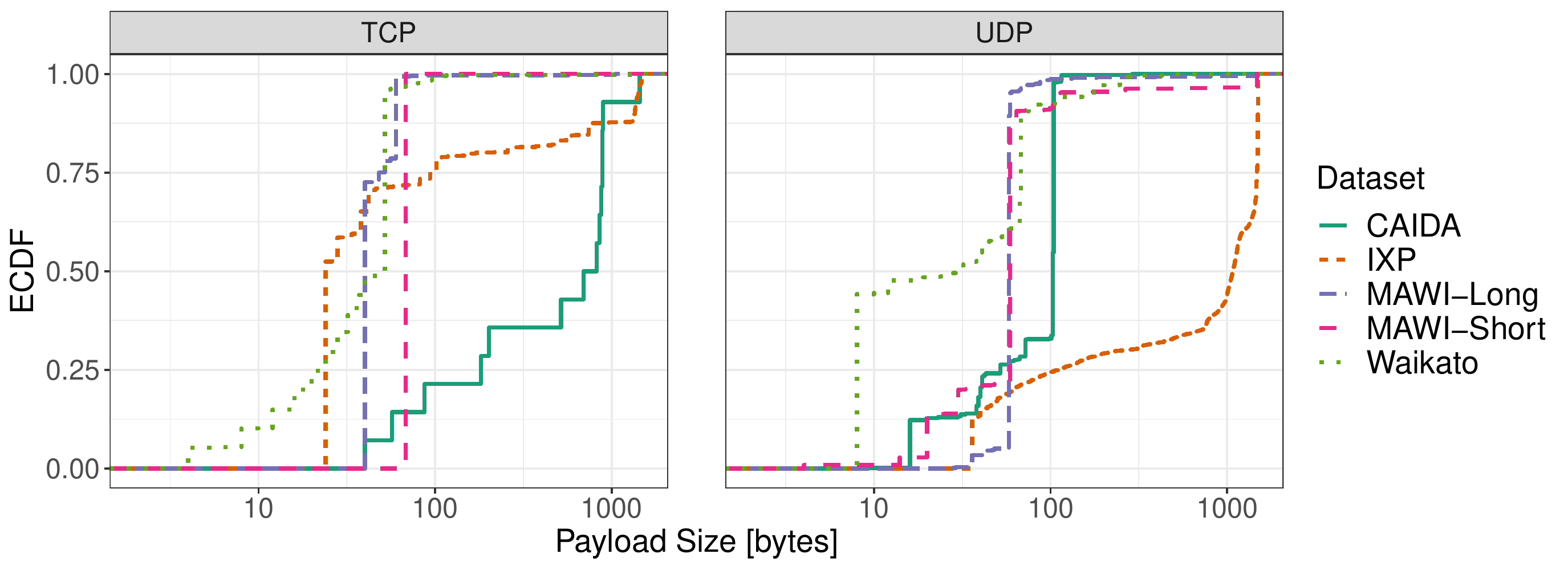}
	\caption{Cumulative distribution of payload size in port 0 traffic. Note that the X-axis is log-scaled.}
	\label{fig:cdf-payload}
\end{figure}

\begin{figure}[b!]
	\captionsetup{skip=.25em}
	\centering
	\includegraphics[width=0.5\columnwidth]{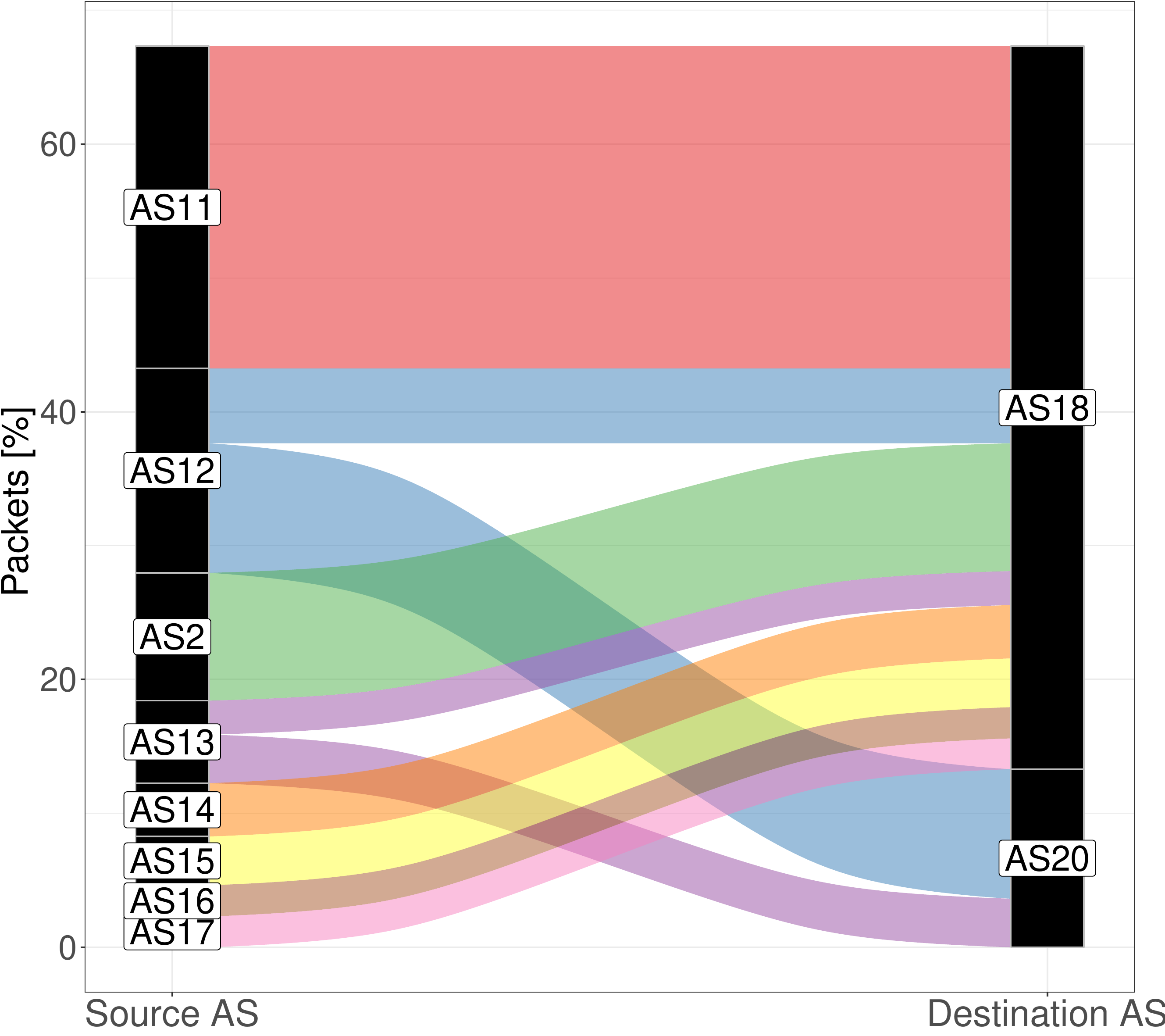}
	\caption{Traffic between top 10 (source AS, destination AS) pairs involved in port 0 traffic in the MAWI-short dataset.}
	\label{fig:alluvial-mawi-asn}
\end{figure}

\begin{figure}[htb]
    \captionsetup{skip=.25em}
    \centering
    \includegraphics[width=\columnwidth]{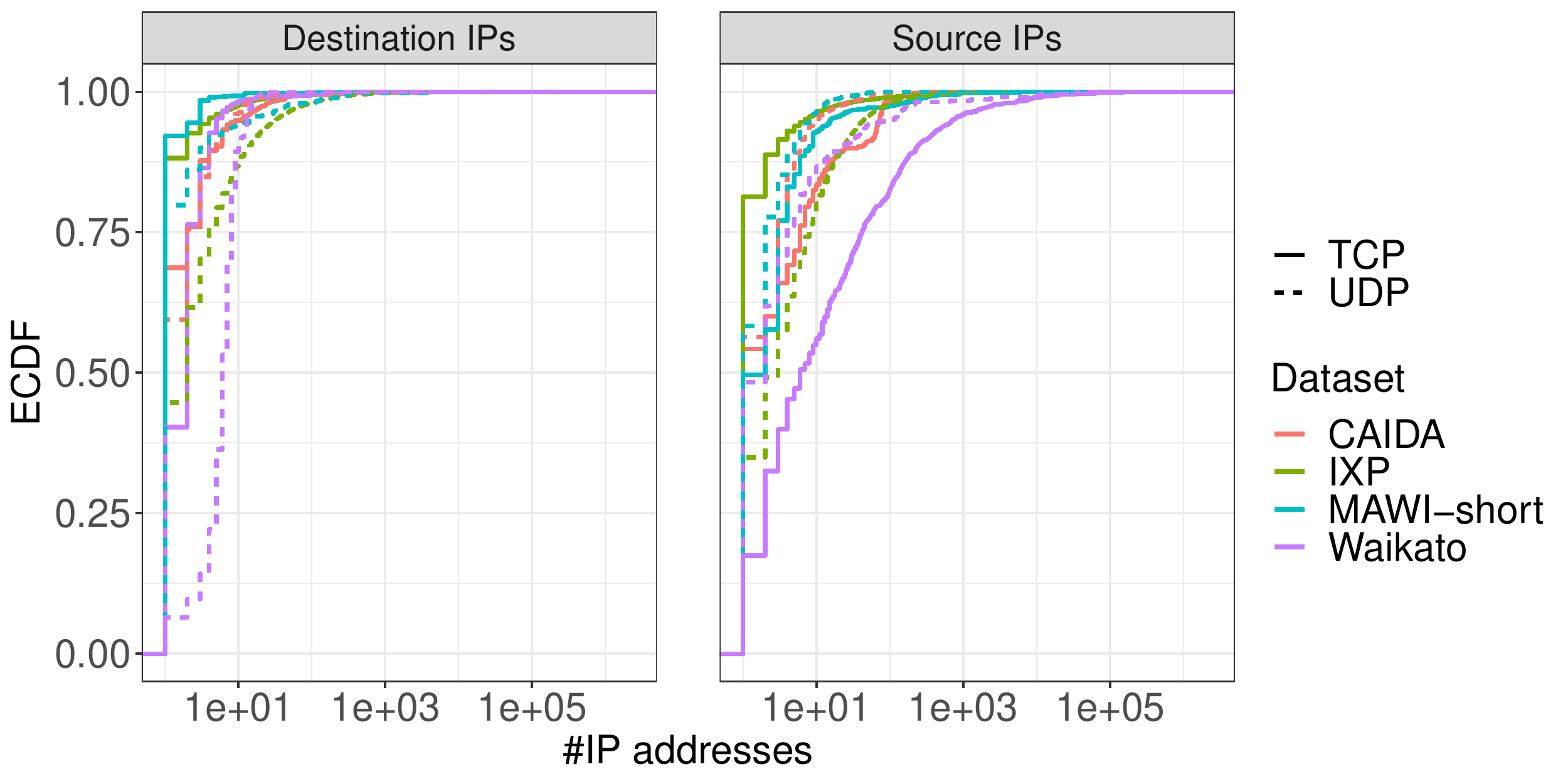}
    \caption{Cumulative distribution of IP addresses in port 0 traffic. Note that the X-axis is log-scaled.}
    \label{fig:cdf-ips}
\end{figure}

Although using a flow-level dataset provides us with useful information about the origin and targets of port 0 traffic, it cannot provide information on what the packets actually contain. 
Knowing the packet content, we can infer the cause of port 0 usage more precisely. 
To this end, we use the MAWI-long, MAWI-short, CAIDA, and Waikato datasets. 
CAIDA contains no payload, while others provide partial payload data.
We begin our packet-level analysis by investigating packet payload sizes,
\new{for which we use the packet length field found in UDP and TCP headers.}
As \Cref{fig:cdf-payload} shows, nearly all packets in MAWI-short, MAWI-long and Waikato have a payload size of less than 100 bytes. 
In both the MAWI-short and the CAIDA dataset, more than 99\% of the TCP port 0 traffic does not have any payload, while UDP traffic always contains payload. 
\new{Note that \Cref{fig:cdf-payload} only shows those TCP packets with payload, \ie for CAIDA and MAWI-short, it shows less than 1\% of all TCP packets.} 
In the CAIDA dataset, while UDP traffic includes payload sizes smaller than 104 bytes in 99\% of the packets, TCP traffic shows more mid-sized payload sizes. 
Investigating further into the CAIDA dataset shows that all packets contain zero as fragment offset and all the fragmentation flags are set to \textit{Don't Fragment}.
\new{This suggests that port 0 traffic in the CAIDA dataset is likely not a fragmentation artifact.}  
However, we find some bogus packets, e.g. with zero header length among these mid-sized TCP packets.
Similar to our analysis in Section \ref{section:flow-level}, we investigate port 0 traffic origins and destinations in the MAWI dataset. 
We find that most of the traffic, namely more than 60\%, is destined to only 2 ASes, as shown in \Cref{fig:alluvial-mawi-asn}.
\Cref{fig:cdf-ips} shows the cumulative distribution of IP addresses in port 0 traffic in different datasets. %
We exclude the MAWI-long dataset since aggregating through 14 years would not give us useful information. 
We observe that more than 75\% of port 0 traffic is originated by less than 10 IP addresses in CAIDA, IXP, and MAWI-short. Also in all the datasets, more than 87\% of port 0 traffic is destined to less than 10 IP addresses.

\begin{figure}
\begin{floatrow}
\ffigbox[6cm]{%
    \includegraphics[width=\columnwidth]{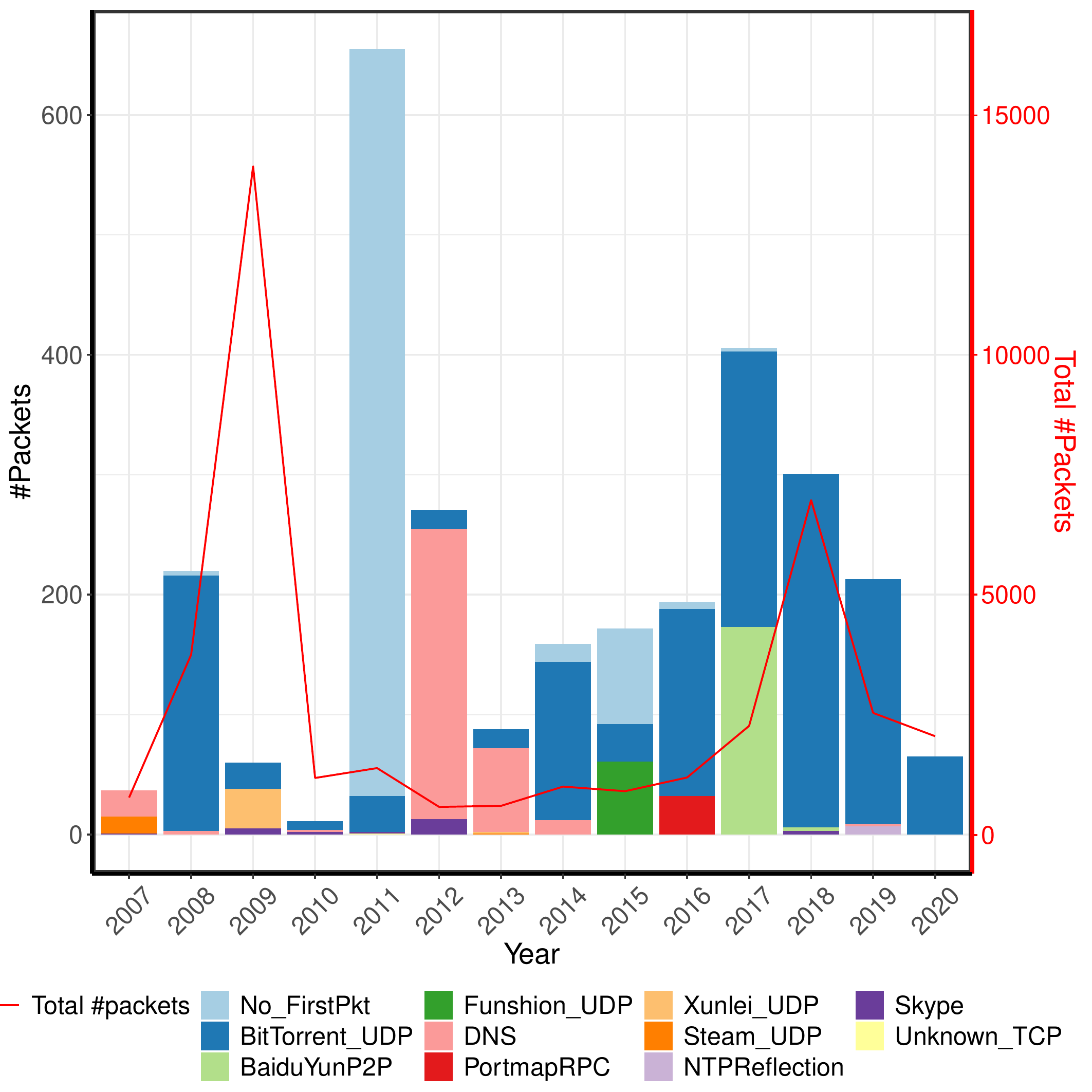}
}{%
\caption{Payload distribution (bar plots) and total packet count (red line) \new{for MAWI-long}.}%
    \label{fig:protoident}%
}

\ffigbox[6cm]{%
    \includegraphics[width=\columnwidth]{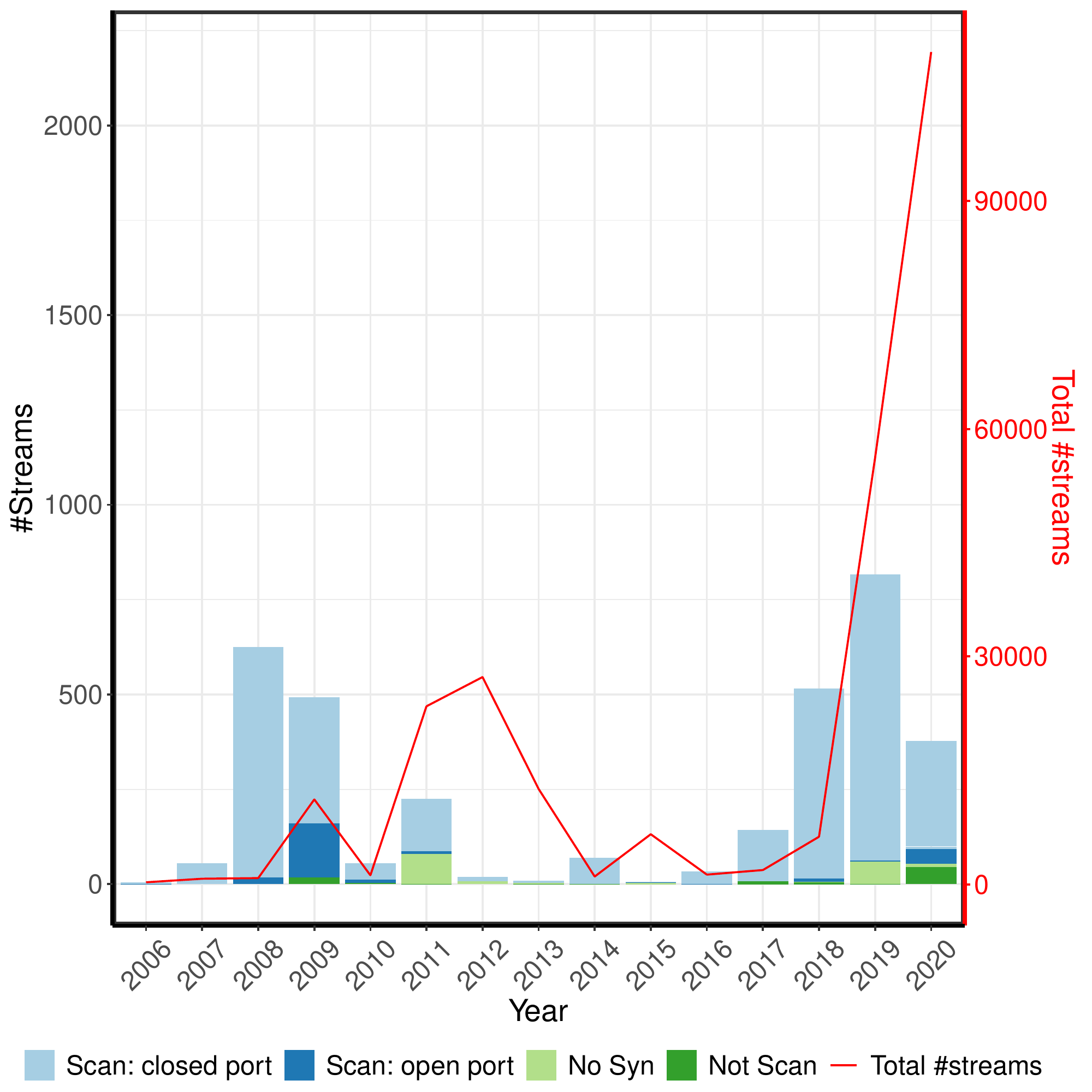}
}{%
\caption{TCP stream categorization (bar plots) and total streams (red line) \new{for MAWI-long}.}%
  \label{fig:tcpstreams}
}
\end{floatrow}
\end{figure}

In \Cref{fig:protoident}, we show the payload distribution classified with libprotoident \cite{alcock2012libprotoident} for each year in the MAWI-long dataset. 
The red line along with the right Y-axis show total number of packets throughout different years. 
The stacked bar plots show different categories of payloads excluding \textit{No Payload} and \textit{Unknown UDP}. 
We find that BitTorrent traffic is a constant contributor to port 0 traffic in Waikato, MAWI-short, and in different years in MAWI-long.

In MAWI-short, we find that 70\% of the payloads belong to the BitTorrent UDP protocol.
Additionally, a payload pattern covering 16\% of the traffic, probably belonging to a custom application-layer protocol, DNS, OpenVPN, and NTP, contributes to other payloads in MAWI-short port 0 traffic in our dataset. 
In Waikato, BitTorrent-UDP and Skype are among the top payloads.

In MAWI-short, MAWI-long, and CAIDA, Malformed packets contribute less than 2\% to port 0 packets, \eg with wrong checksums, having UDP length of higher than IP length, etc. 
However, in Waikato, we find that 16.2\% of the traffic is malformed. 
This shows that port 0 traffic can also be caused by misconfiguration, programming errors, or people sending malformed traffic on purpose.
Next, we analyze different TCP flags in packet-level datasets to better understand possible causes of port 0 traffic.
Attackers and scanners usually use specific TCP control bits in their packets to achieve their goals. 
For instance, attackers sending spoofed traffic set the SYN bit to try to initiate TCP connections with their targets, which in backscatter traffic we see as SYN/ACK, RST, RST/ACK, or ACK packets \cite{wustrow2010backscatter}. Therefore, we investigate TCP control flags in the datasets. We observe that most of the TCP flags are only SYNs: More than 66\% in MAWI-short, and 92\% in CAIDA, which might indicate that most of the TCP port 0 traffic in these two datasets is caused by scanning.
We analyze TCP flags in MAWI-long dataset per year, as shown in \Cref{fig:tcpstreams}.
First, we check whether all packets in a TCP stream are one-way or two-way.
We find that a large fraction of the TCP streams are one-way. This also holds for all other packet-level datasets.
Then, we categorize two-way TCP streams as follows:
\begin{itemize}
\item Scan to closed port: Client sends SYN, receives RST or RST/ACK.
\item Scan to open port: Client sends SYN, receives SYN/ACK, client then sends RST or RST/ACK.
\item No SYN: No SYN is ever sent. \new{The stream begins with other flags, mostly SYN/ACKs followed by RSTs from the other side.}%
\item Not scan: None of the above, \new{\ie client sends SYN but receives no RST.}
\end{itemize}
We find that a major fraction of two-way TCP streams are scans to closed ports for most of the years.
\new{Among the streams in the \emph{Not scan} category, we find two long streams of ACK/PSH followed by multiple ACKs in 2015 or ACK/PSH/FIN in 2019, respectively. 
We believe that these streams are related to an ACK/PSH flood attack \cite{hallman2017ioddos} considering the relatively high number of packets sent in these streams.}
\todoam{Please check if the above rephrasing correctly captures what's going on.}
Next, we analyze specific years of the MAWI-long dataset with very characteristic spikes more in-depth.
In 2009, we see the largest number of total packets of any year, with a TCP:UDP ratio of about 2:1.
The majority of UDP traffic is originating with source port UDP/8000 from many different IP addresses within a Chinese ISP AS which are mostly destined to UDP/0 towards a single IP address belonging to a Japanese university inside WIDE.
For TCP, the majority of traffic is sourced from a single IP address within a Canadian ISP and destined to many different IP addresses.
Almost all sources are TCP/0 and the destinations are TCP/22 (SSH).
As is shown in \Cref{fig:tcpstreams}, these are very likely scanning activities.

In 2012 we see the largest number of TCP streams as shown in \Cref{fig:tcpstreams}.
We find a factor of 54 times more TCP traffic this year than UDP traffic.
Almost \sperc{80} of all TCP/0 traffic is from a single IP address within a hosting company, the destination addresses and ports are evenly distributed.
The TCP flags of all packets are set to RST/ACK.
These indicators lead us to believe that this is backscatter traffic from attack traffic using spoofed IP addresses \cite{wustrow2010backscatter}.

Finally, we investigate the current year 2020, from January to July.
During this period we see 26 times as much TCP traffic compared to UDP.
The majority of TCP traffic originates from a single IP address at a hosting company, which uses TCP/43573 as a source port.
For the IP address in question we find many different reports on abuse DB websites, which hint at scanning and vulnerability probing.

To summarize, we find that a large fraction of TCP streams in port 0 traffic is one-way. 
However, we still see some two-way streams related to scanning activities.
Analyzing packet payloads throughout all our datasets, we observe that BitTorrent UDP traffic is a constant contributor to port 0 traffic.
\section{Active Measurements} \label{section:active-measurement}
\budget{1.25}

As discussed in the previous sections, we observed a significant number of RST/ACKs and even some SYN/ACKs which indicate scanning activities.
To better understand how the network reacts to port 0 traffic, we stage an active measurement campaign.
We run two types of measurements: \one Port scan measurements allow us to analyze responsiveness of IP addresses to port 0 probes and \two traceroute measurements provide information on where port 0 packets are being filtered.

\subsection{Responsive Addresses}

We run four types of port scan measurements, for each possible combination of IPv4/IPv6 and TCP/UDP.
The IPv4 measurements are run on the complete address space minus a blocklist, the IPv6 measurements use an IPv6 hitlist \cite{hlv2}.
For the TCP measurements we send regular SYN packets, for UDP we send the most prominent payload found in our passive packet traces.

\begin{figure}
\begin{floatrow}
    \ffigbox[7cm]{%
    \includegraphics[width=\columnwidth]{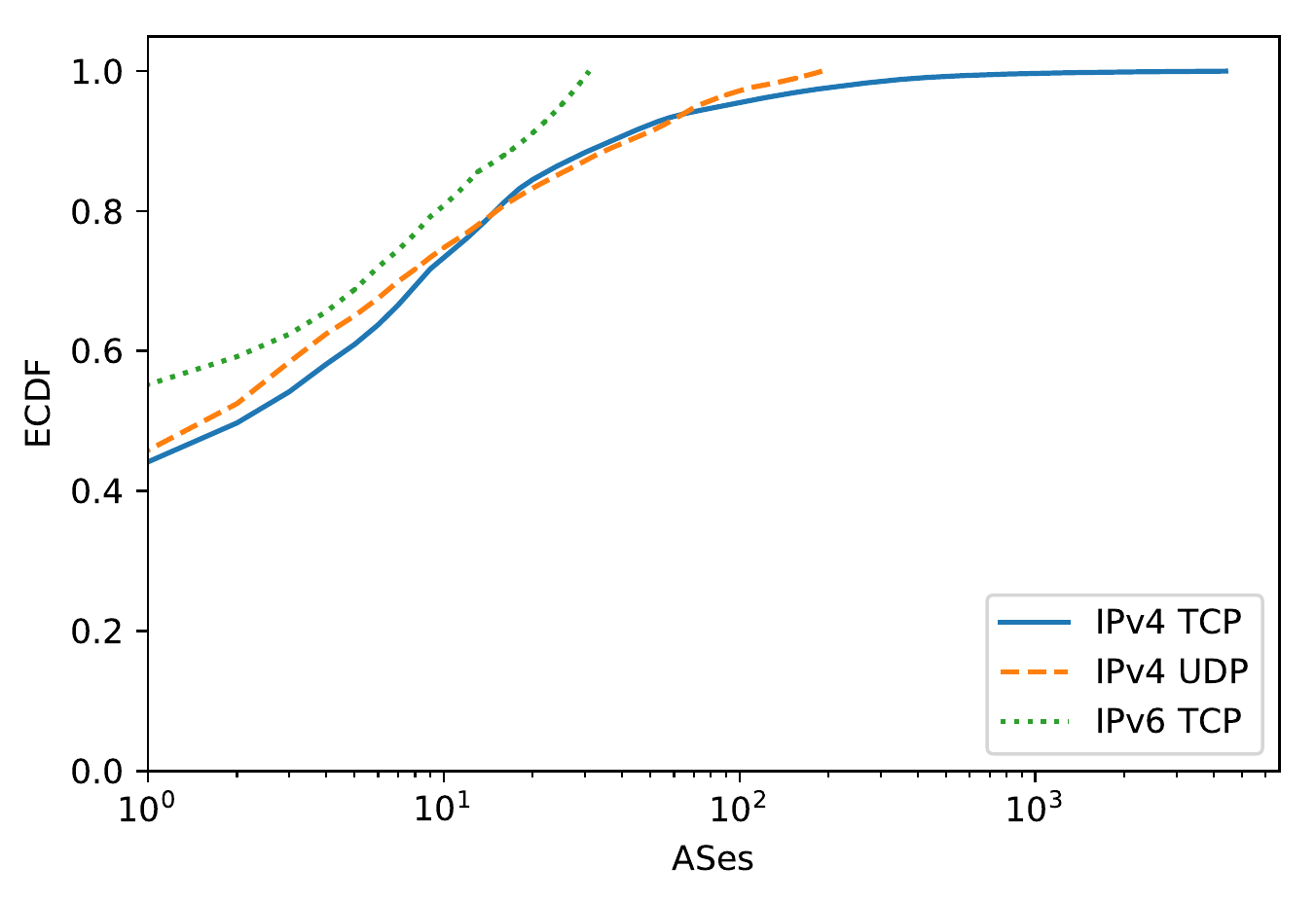}
}{%
    \caption{Cumulative distribution of responsive IP addresses per AS. Note that the X-axis is log-scaled.}%
    \label{fig:active-measurements-as-cdf}%
}
\capbtabbox[5cm]{%
    \begin{tabular}{lrlr}
\toprule
{} &    ASN &           AS Name &  Count \\
\midrule
1  &   6830 &    Liberty Global &   4822 \\
2  &   6327 &              Shaw &   3257 \\
3  &    812 &            Rogers &   2297 \\
4  &  33915 &          Vodafone &   2152 \\
5  &  11492 &         Cable One &   1095 \\
6  &  30036 &          Mediacom &    688 \\
7  &  12389 &        Rostelecom &    643 \\
8  &   4134 &          Chinanet &    575 \\
9  &   3320 &  Deutsche Telekom &    552 \\
10 &   4766 &     Korea Telecom &    498 \\
\bottomrule
\end{tabular}
}{%
  \caption{Top 10 ASes of non-reachable target addresses when comparing TCP/0 and TCP/80.}%
  \label{tab:top-ases-traceroutes}
}
\end{floatrow}
\end{figure}

For the four protocol combinations, we get vastly differing results.
With \sm{2.3}, the largest number of addresses responds to our IPv4 TCP port 0 probes.
Only 2222 unique addresses respond to IPv4 UDP probes and 120 respond to IPv6 TCP probes.
We find not a single responsive address for IPv6 UDP probes.

When mapping responsive addresses to ASes \cite{pyasn,routeviews}, we find that a small number of ASes makes up the majority of responses.
\Cref{fig:active-measurements-as-cdf} shows the AS distribution of responses for the different protocols.
The top ten ASes make up \sperc{72}, \sperc{73}, and \sperc{79} of all responses for IPv4 TCP, IPv4 UDP, and IPv6 TCP, respectively.
When we look at the overlap of responding addresses in TCP and UDP for IPv4, we find that \sperc{61} of IPv4 UDP addresses are present in IPv4 TCP results.
In IPv4 TCP, where we see the most responses by far, most of the top 10 ASes belong to ISPs.
This leads us to believe that faulty or misconfigured ISP equipment is to blame for responses to port 0 probes.

Next, we analyze the initial TTL (iTTL) value \cite{mukaddam2014ip,jin2003hop,backes2016feasibility}, UDP reply payload, and combine these with the responding AS.
For IPv4 TCP we find that the most common iTTL values are 64 (\sperc{57}), which is the default for Linux and macOS, 255 (\sperc{36}), the default for many Unix devices, and 128 (\sperc{7}) the default for Windows.
When combining these iTTL values with the responding AS we find no clear patterns.
In contrast, for IPv4 UDP we find a clear correlation between iTTL, payload, and AS.
The most common response payload (\sperc{32}) is sent from six different ASes with an iTTL value of 32 or 64.
The second most common response payload (\sperc{14}) is identical to our request payload, \ie the probed hosts simply mirror the payload that they receive.
Packets with this payload originate from a single AS (AS7922, Comcast) and all of them have an iTTL of 255.
The third most common payload (\sperc{8}) is made up of 16 zero bytes and originates from AS14745 (Internap Corporation) with an iTTL of 32.

These findings suggest that only a small number of networks contain misconfigured devices erroneously responding to port 0 probes.

\subsection{Port 0 Traceroutes}

To better understand how port 0 traffic is handled inside the network, we conduct traceroute-style measurements using \yarrp~\cite{yarrp}.
This allows us to see if port 0 traffic is treated differently by routers compared to standard TCP/80 or TCP/443 traffic.\footnote{Note that due to the nature of traceroute measurements, missing traceroute responses could stem either from filtered packets on the forward path, rate-limiting of ICMP packets at the routers, as well as dropping of ICMP responses on the return path.}
In IPv4, we split the announced address space into \sm{11} /24 prefixes and send a trace to a random address within each of these prefixes.
In IPv6, the equivalent would be sending traces to every /48 prefix.
This is, however, not feasible due to the vast address space.
Therefore we decide to pick one random address per announced IPv6 prefix, no matter the prefix length.
We ensure that random addresses for less specific prefixes do not fall into more specific prefixes.
In total, we send probes to about \sk{88} IPv6 prefixes.

When analyzing the reached target addresses depending on the used port numbers, we find that in IPv4 there is a significant difference between port 0 and other ports.
\sk{91} of IPv4 port 0 traces reach their target, whereas \sk{118} traces on TCP/80 and TCP/443 reach their target IPv4 address, an increase of almost \sperc{30}.

In IPv6, however, almost no targets are reached for either port number, as the likelihood of a randomly generated address in a prefix actually being assigned is quite low.
Therefore, we perform additional analyses based on the reachability of the target BGP-announced prefix.

The general picture in IPv4 does not change drastically when analyzing the reachability of the target prefix:
Port 0 probes reach fewer target prefixes compared to port 80 and port 443 probes, although the difference is reduced to \sperc{14.2} and \sperc{9.5}, respectively.

When we analyze the reached target prefixes for IPv6, however, we see a slight difference of \sperc{3}.

As the difference of reachable addresses is most apparent in IPv4, we investigate this phenomenon in more detail.
We identify on a per-target basis the addresses which see no responses in TCP/0, but do see responses in TCP/80.
These non-responsive port 0 addresses are mapped to 4102 distinct ASes, exhibiting a long-tailed distribution.
Next, we check whether we find other addresses in these 4102 ASes to be responsive to port 0 traceroutes, to exclude the possibility of missing responses due to ICMP rate limiting.
We find responses to port 0 traceroutes for only 15 of these ASes, making up only \sperc{0.4} of the total 4102 ASes.
This underlines the fact that these ASes are indeed handling port 0 traceroutes differently compared to other ports.
Furthermore, as is shown in \Cref{tab:top-ases-traceroutes}, 9 out of the top 10 ASes belong to ISPs, further indicating that these might be ASes blocking port 0 traffic to their clients\cite{nanog,xfinity,att}.
We analyzed many additional aspects of traceroute responses, by checking for differences in the last responsive hop, comparing the number of responsive hops per trace, evaluating ICMP types and codes, but finding no additional differences between traceroutes using port 0 compared to other ports.
We provide these results for the interested reader in \Cref{sec:appendix:traceroute}.

To summarize, our findings show that packets are handled differently based on the destination port number.
\new{Port 0 is more likely to be filtered on the path as well as at the target hosts.}
Interestingly, the phenomenon of fewer responses for TCP/0 seems to be much more common in IPv4 compared to IPv6, which could be due to inconsistent firewall rules \cite{czyz2016don}.
\section{Conclusion} \label{section:conclusion}
\budget{0.25}

In this work, we dissected port 0 traffic by analyzing five complementing passive datasets and by conducting active measurements.
We showed that the majority of port 0 traffic in the wild flows between a small number of source and destination ASes/IP addresses.
Moreover, for some ASes we identified similar diurnal patterns in port 0 traffic as with regular traffic, along with many TCP packets with no TCP flags, hinting at a prevalence of fragmented traffic in the IXP dataset. 
Additionally, we found that a major fraction of UDP port 0 traffic contains payload, with BitTorrent being a common contributor.
Moreover, we showed that TCP port 0 traffic usually does not contain any payload and is mostly one-way.
Two-way streams were identified as mostly scanning traffic.
Finally, by staging an active measurement campaign, we showed unusually high response rates to TCP port 0 probes in IPv4, in addition to uncovering the presence of port 0 packet filtering.

\noindent\new{\noindent\textbf{Acknowledgments:}
We are thankful to the anonymous reviwers as well as our shepherd Ramakrishna Padmanabhan for their constructive feedback.
We also thank the \ixp, MAWI, the University of Waikato, and CAIDA for providing the data used in our analysis.}

\balance

\bibliographystyle{splncs04}
\bibliography{paper}

\clearpage
\appendix

\section{Additional Traceroute Analyses}
\label{sec:appendix:traceroute}

We perform additional analyses for the active traceroute measurements, which we provide in the following.

\subsection{Last Responsive Hops}

We analyze the last responsive hop of each trace specifically.
More concretely, we are interested in the distance, \ie the largest TTL value of traceroutes, where we get an ICMP response to.
This allows us to determine whether TCP/0 traceroutes are \eg dropped earlier in the network and therefore are terminated earlier in the Internet.

\begin{figure}
    \includegraphics[width=.5\columnwidth]{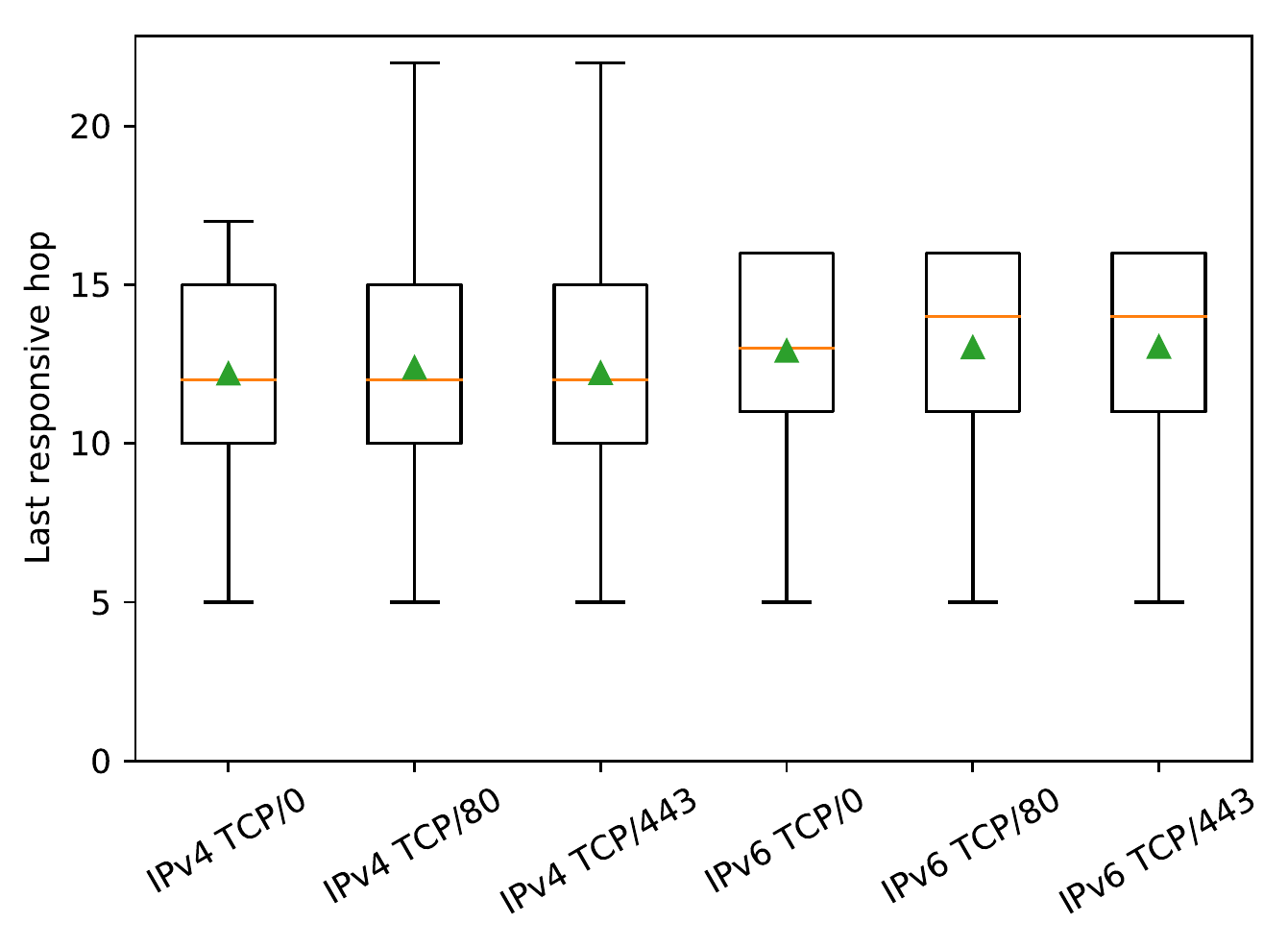}%
    \includegraphics[width=.5\columnwidth]{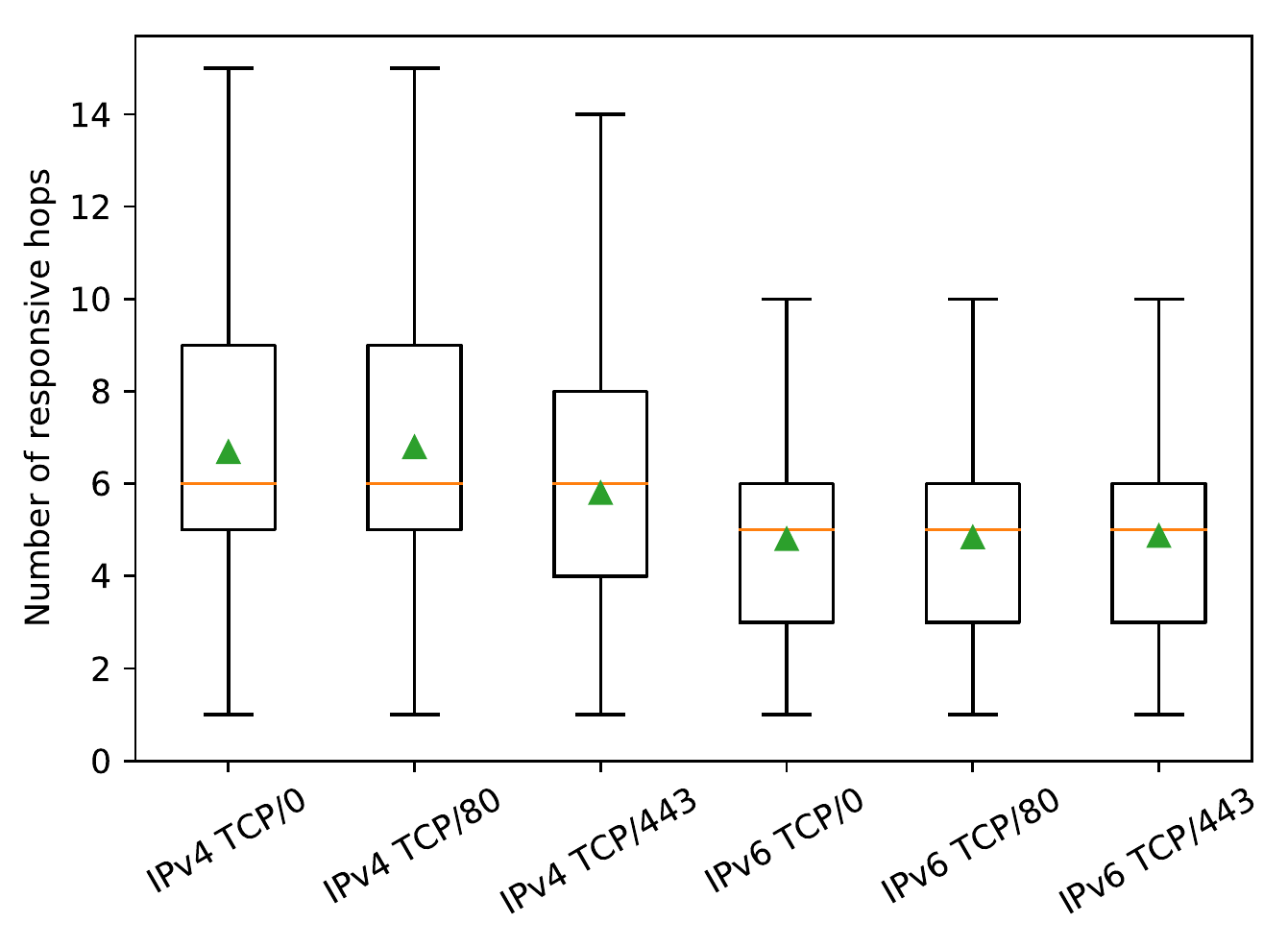}%
    \caption{Box plot of last responsive hop (left) and the number of responsive hops (right) aggregated by transport port protocol for IPv4 and IPv6 showing the median, first and third quantiles, mean ($\blacktriangle$), and 1.5 times IQR as whiskers.}
    \label{fig:yarrp-hops}
\end{figure}

Therefore, we compare the distribution of the last responsive hop.
The left part of \Cref{fig:yarrp-hops} shows the distribution of the last responsive hop for IPv4 and IPv6, respectively.
The only visible difference we see for IPv4 are the lower whiskers for TCP/0, stemming from the fact that TCP/80 and TCP/443 has slightly more outliers with high TTLs when it comes to the last responsive hops.
For IPv6 we see that TCP/0 has a median of 13 and TCP/80 as well as TCP/443 have a median last responsive hop TTL of 14.
Since the median is almost identical, this is due to the median only being able to represent integer values if all elements (namely path lengths) are integers.
TCP/0's median is therefore ``just below'' 14 and the others' median is ``just above'' 14.
All in all, the box plots show that there is no significant difference when analyzing last responsive hops depending on the transport port.

\subsection{Number of Responsive Hops}

Next, we try to answer the question whether fewer routers on the path send ICMP messages for port 0 traceroute traffic or not.

In the right part of \Cref{fig:yarrp-hops} we show the box plot of the number of responsive hops.
Again, we see no evidence of router sending fewer ICMP responses for port 0 traffic.
We see a slight reduction of TCP/443 ICMP responses per trace in IPv4.

\subsection{ICMP Types and Codes}

Finally, we evaluate the different ICMP types and codes sent by routers.

\begin{figure}
    \includegraphics[width=.5\columnwidth]{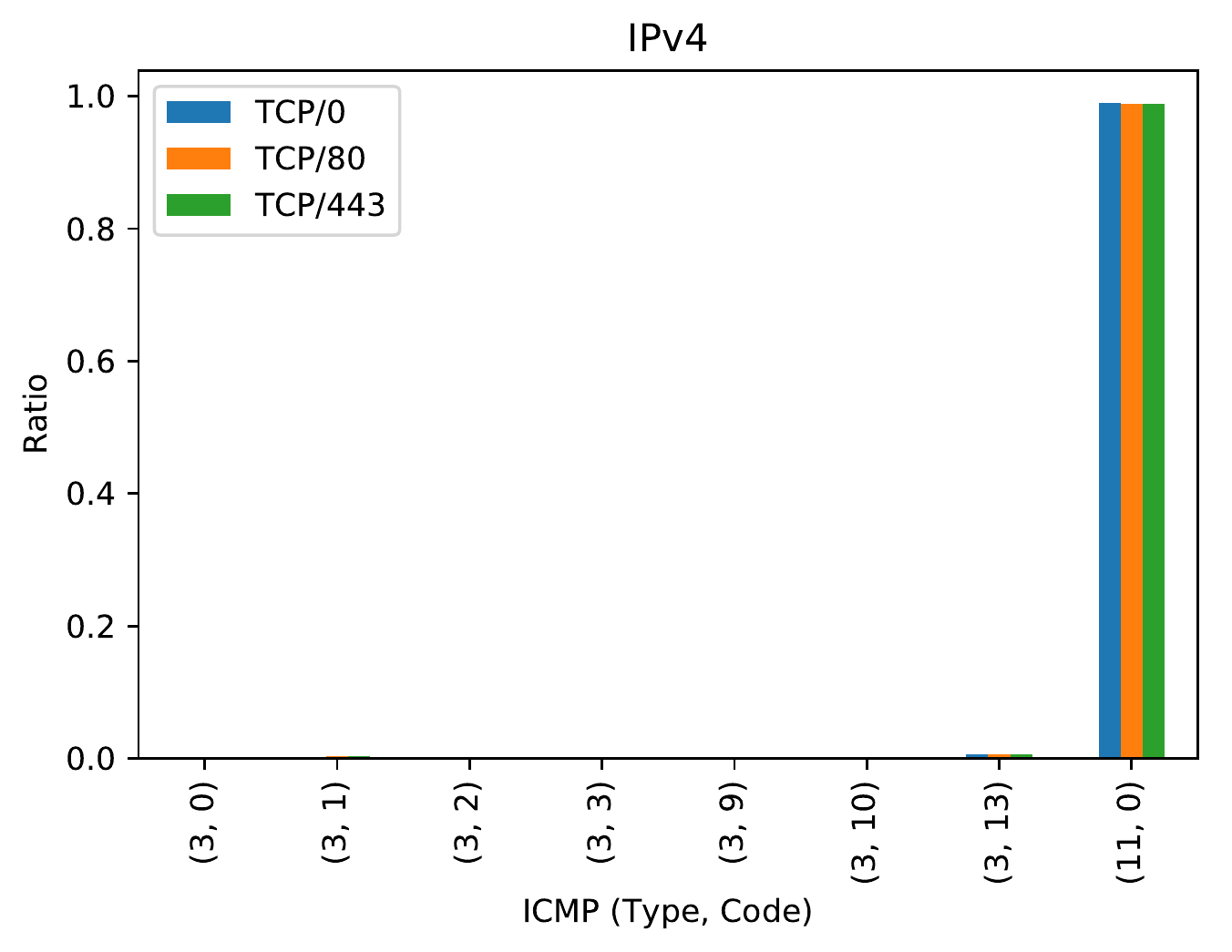}%
    \includegraphics[width=.5\columnwidth]{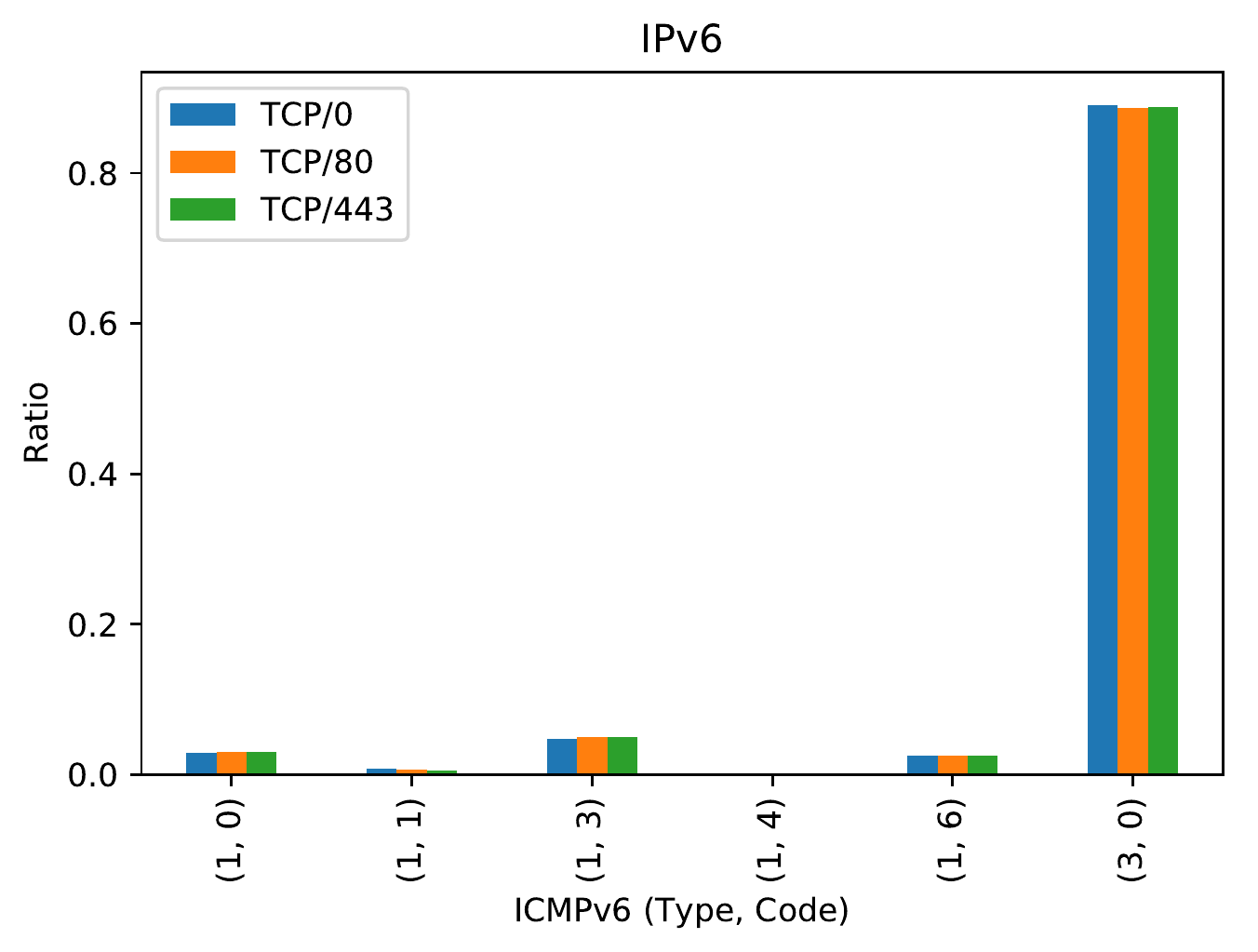}%
    \caption{Distribution of ICMP(v6) type and code combinations for all responses split by transport protocol for IPv4 (left) and IPv6 (right).}
    \label{fig:yarrp-icmp-type-code}
\end{figure}

\Cref{fig:yarrp-icmp-type-code} shows the distribution of type and code combinations for ICMP and ICMPv6, respectively.
As expected, the vast majority are of type ``Time to Live exceeded in Transit'' for IPv4 and `hop limit exceeded in transit'' for IPv6.
We see almost identical distributions for the port 0 and other ports.

\section{Additional Passive Analysis}
\label{sec:appendix:passive}

\new{We analyze hourly patterns of port 0 traffic grouped by source AS, compared with the total port 80 traffic as a reference for regular traffic.
Due to space limitations we publish the figure on our website:}

\begin{center}
    \new{\href{https://inet-port0.mpi-inf.mpg.de/}{\textbf{inet-port0.mpi-inf.mpg.de}}}
\end{center}

 \clearpage

\end{document}